\documentclass[prl,aps,twocolumn,showpacs,preprintnumbers,amsmath,amssymb]{revtex4}
\usepackage{epsfig}
\usepackage{color}
\usepackage{ifthen}
\usepackage{xfrac}
\usepackage{graphicx}
\usepackage{amsmath}
\usepackage{amssymb}

\usepackage{graphicx}% Include figure files
\usepackage{dcolumn}% Align table columns on decimal point
\usepackage{bm}% bold math
\def\Dsl{\hbox{/\kern-.6700em\it D}} % D slash
\def\dsl{\hbox{/\kern-.5300em$\partial$}}

\def\eqa{\begin{eqnarray}}
\def\eeqa{\end{eqnarray}}
\def\eq{\begin{equation}}
\def\eeq{\end{equation}}
\def\be{\begin{equation}}
\def\ee{\end{equation}}
\def\bea{\begin{eqnarray}}
\def\eea{\end{eqnarray}}

\newcommand{\dslash}{\not{\hbox{\kern-2pt $\partial$}}}
\newcommand{\pslash}{\not{\hbox{\kern-2.3pt $p$}}}
 \newtoks\nslashfraction
 \nslashfraction={.13}
 \newcommand{\nslash}[1]{\setbox0\hbox{$ #1 $}
   \setbox0\hbox to \the\nslashfraction\wd0{\hss \box0}/\box0 }

\begin{document}

\preprint{}

\title{Model Independent Signatures of New Physics in the Inflationary Power Spectrum}
\author{Mark G. Jackson and Koenraad Schalm}
\affiliation{Instituut-Lorentz for Theoretical Physics, University of Leiden, Leiden 2333CA, The Netherlands}

\date{\today}

\pacs{04.62.+v, 98.80.-k, 98.70.Vc}

\begin{abstract}
\noindent
We compute the universal generic corrections to the inflationary power spectrum due to unknown high-energy physics. We arrive at this result via a careful integrating out of massive fields in the ``in-in'' formalism yielding a consistent and predictive low-energy effective description in time-dependent backgrounds. We find that the power spectrum is universally modified at order $H/M$, where $H$ is the scale of inflation.  This is qualitatively different from the universal corrections in time-independent backgrounds, and it suggests that such effects may be present in upcoming cosmological observations.
\end{abstract}

\maketitle

%%%%%%%%%%%%%%%%%%%%%%%%
\subsection{Introduction}
%%%%%%%%%%%%%%%%%%%%%%%%

Inflationary theory 
has become a cornerstone of modern cosmology, elegantly solving many problems with Standard Big Bang Cosmology and predicting the primordial power spectrum whose evolution determines the temperature fluctuations in Cosmic Microwave Background.  These are now matched to observation with spectacular precision \cite{wmap7}.  Together with the realization that the inflationary energy scale may not be far from that of quantum gravity, the continuing advance in high precision observation may provide an opportunity to observe new fundamental physics near the Planck scale of quantum gravity in cosmological data \cite{Brandenberger:1999sw,Bergstrom:2002yd,Martin:2003sg,Martin:2004iv,Martin:2004yi,Easther:2004vq,Greene:2005aj,Easther:2005yr,Spergel:2006hy}. In this letter we shall show and compute potentially observable {\em universal generic} corrections to the prediction of inflation that are {\em independent} of the precise details of the theory of quantum gravity or other unknown physics near the Planck scale.

Any new fundamental physics signals are small corrections to the existing measured characteristics and can therefore only be seen if the effects are large enough that they can be detected with upcoming precision experiments such as \emph{Planck} \cite{planck} or \emph{CMBPol/Inflation Probe} \cite{cmbpol}.  The primary measurement of interest is the primordial density (scalar) fluctuation power spectrum $P_{s}(k)$ itself.  This has been observationally determined \cite{wmap7} to be very nearly scale-invariant:
\begin{equation}
\label{pwmap5}
 P_{s}(k)   \sim k^{n_s-1}, \hspace{0.5in} n_s \approx 0.960 \pm 0.013 . 
\end{equation}
Inflationary theories predict the amplitude and the momentum dependence, in particular the value of $n_s$.  % Despite the importance of (\ref{pwmap5}) in constraining theoretical models of the early Universe, its computation suffers from an inconsistency that for most models of inflation one is formally working in an energy regime far beyond the Planck-scale, where gravitational backreaction and quantum-gravity corrections cannot be ignored \cite{Brandenberger:1999sw}. This `Transplanckian problem' is really a window of opportunity: the non-negligible nature of these corrections implies that New Physics arising at high energy scales could have a measurable effect on the power spectrum.
The question of the observability of Planck scale corrections to $P_s(k)$ was actively pursued some time ago with the conclusion that in toy models \cite{Niemeyer:2000eh,Kempf:2000ac,Niemeyer:2001qe,Kempf:2001fa,Martin:2000xs,Brandenberger:2000wr,Brandenberger:2002hs,Martin:2003kp,Easther:2001fi,Easther:2001fz,Easther:2002xe,Kaloper:2002uj,Kaloper:2002cs,Danielsson:2002kx,Danielsson:2002qh,Shankaranarayanan:2002ax,Hassan:2002qk,Goldstein:2002fc,Bozza:2003pr,Alberghi:2003am,Schalm:2004qk,porrati,Porrati:2004dm,Hamann:2008yx}
one can obtain measurable corrections of the order $H/M$, comparable to intrinsic cosmic variance, with $H \lesssim 10^{14}$ GeV the Hubble scale during inflation and $M$ the energy scale of New Physics.  It is believed that such corrections linear in $H/M$ encode fundamental physics effects on the initial state rather than the dynamics.  Broadly put, all previously considered models fall into two classes:
\begin{itemize}
\item[(a)] New Physics Hypersurface (NPH) models, where initial conditions for each momentum mode are set at the redshift where it equals the scale of New Physics.  All such initial conditions are ad hoc and lack a direct connection with the New Physics. 
\item[(b)] Boundary Effective Field Theory (BEFT) models, which have a manifest connection with the New Physics. This framework is not universal as the effects are controlled by the initial time of inflation, rather than the redshift where New Physics becomes relevant. 
\end{itemize}
%KS Start
To make a definitive statement,
%KS End 
one needs the {\em universal generic} model-independent corrections to the power spectrum in terms of an effective field theory (provided adiabaticity is maintained \cite{Burgess:2002ub,Burgess:2003zw,Burgess:2003hw}).  This long-standing question has been hampered by the obstacle of constructing low energy effective theories in cosmological spacetimes where energy is not a conserved quantity.

Here we use our recent insight on how this obstacle can be overcome to compute the universal generic new physics corrections to the inflationary power spectrum.  One can generate the universal low energy effective action by integrating out a massive field in any particular New Physics model.  This is sensible in a cosmological setting, as long as one computes expectation values directly, 
%KS Start
rather than transition amplitudes. 
%KS End
The details behind the construction of low-energy effective actions in cosmological backgrounds will be given in a companion article \cite{jacksonschalmvdaalst}. 
%%%%%%%%%%%%%%%%%%%%%%%%%%%%
\subsection{Universal Corrections to the Power Spectrum}
%%%%%%%%%%%%%%%%%%%%%%%%%%%%
To demonstrate how this procedure works in practice, let us consider an example of New Physics.  The simplest theories of inflation are a single scalar field $\phi$ coupled to gravity,
\begin{equation}
\label{sinf}
 S_{\inf} [\phi] = \int d^4 x \sqrt{g} \left[\frac{1}{2} M^2_{\rm pl} R- \frac{1}{2} (\partial \phi)^2 - V(\phi) \right] . 
\end{equation}
The fluctuations $\varphi(t,{\bf x}) \equiv \delta \phi$ around a classical background solution $\phi_0(t)$ that inflates determine the spectrum (\ref{pwmap5}).  This power spectrum is computed through the equal-time two-point correlation function via the `in-in' formalism \cite{Calzetta:1986cq}:
\begin{equation}
\label{ps}
P_{\varphi} (k) \equiv \lim_{t \rightarrow \infty} \frac{k^3}{2\pi^2} \langle {\rm in}(t) | \varphi_{\bf k} (t) \varphi_{-\bf k} (t) | {\rm in}(t) \rangle.
\end{equation}
Traditionally, the in-state $| {\rm in} \rangle$ is taken to be the Bunch-Davies vacuum state, but this is not necessarily so.  Expanding cosmological backgrounds allow for a more general class of vacua, which can be heuristically considered to be excited states of inflaton fluctuations.  In the present context, we will find that integrating out high-energy physics generically results in boundary terms in the effective action, which represent such excited states. This gives a qualitative connection with the aforementioned toy-models with potentially observable corrections.

To the inflationary action (\ref{sinf}) we add a massive field $\chi$ with a simple interaction to the inflaton fluctuations,
\begin{eqnarray*}
\label{action}
S_{\rm new}[\varphi,\chi] &=& - \int d^4 x \sqrt{g} \left[  \frac{1}{2} (\partial \chi)^2 + \frac{1}{2} M^2 \chi^2 + \frac{g}{2} \varphi^2 \chi \right].
 \end{eqnarray*}
We ignore self-interactions of $\varphi$ because we are only interested in scale-dependent corrections.  There are no linear terms in the fluctuations, ensuring that a solution to the action $S_{\rm inf}$ is also a solution of the combined action $S \equiv S_{\rm inf} + S_{\rm new}$.  If $V(\phi)$ has a minimum at a value $\phi_0$ with $V(\phi_0)>0$, the combined action $S$  produces an inflationary phase de Sitter background metric
\begin{eqnarray} 
\label{eq:2}
ds^2 = a(\tau)^2 (-d \tau^2 + d {\bf x}^2), \hspace{0.4in} a(\tau) = - 1/H \tau 
\end{eqnarray}
with a constant Hubble scale $H$, 
but contains new physics in the fluctuations parameterized by $g$ and $M$. As de Sitter inflation is representative for all slow-roll models, we shall take  \eqref{eq:2} as the metric background.  

For non-equilibrium systems such as a cosmological background the fundamentally sound approach to computing expectation values such as the power spectrum (\ref{ps}) is the Schwinger-Keldysh approach.  At some early time $t_{\rm in}$ we impose the Bunch-Davies vacuum $| {\bf 0} \rangle$ for $\varphi$ and $\chi$, evolve the system for the bra- and ket-state separately until some late time $t$, and then evaluate the two-point fluctuation correlation:
\begin{eqnarray} \label{correlation} P_{\varphi}(k) &=& \lim_{t \rightarrow \infty} \frac{k^3}{2\pi^2} \times \\
\nonumber
&& \hspace{-0.5in} \langle {\bf 0}(t_{\rm in}) | e^{i \int_{t_{\rm in}}^t dt' \mathcal H (t')} | \varphi_{\bf k} (t)|^2 e^{-i \int_{t_{\rm in}}^t dt'' \mathcal H(t'')} | {\bf 0}(t_{\rm in}) \rangle. 
 \end{eqnarray}
Focusing now on the fluctuations in the action, if we denote the fields representing the ``evolving'' ket to be $\{ \varphi_+, \chi_+ \}$ and those for the ``devolving'' bra to be $\{ \varphi_-, \chi_- \}$, the in-in expectation value (\ref{correlation}) can be computed from a path-integral with action
\[ \mathcal S \equiv S[\varphi_+, \chi_+] - S[\varphi_-, \chi_-]  \]
together with the constraint that $\varphi_-(t)=\varphi_+(t)$ and $\chi_-(t)=\chi_+(t)$.
It is then helpful to transform into the Keldysh basis,
\begin{eqnarray*}\
\nonumber
 {\bar \varphi} &\equiv& (\varphi_+ + \varphi_-)/2,  \hspace{0.5in} {\Phi} \equiv \varphi_+ - \varphi_-, \\
{\bar \chi} &\equiv& (\chi_+ + \chi_-)/2,  \hspace{0.5in} {\rm X} \equiv \chi_+ - \chi_-
\end{eqnarray*}
where the action equals
\begin{eqnarray}
\label{eq:1}
\nonumber
&& \hspace{-0.2in} \mathcal S [ {\bar \varphi}, {\Phi}, {\bar \chi},  {\rm X}] = - \int d^4 x \sqrt{g} \left[ \partial {\bar \varphi} \partial \Phi  +  \partial {\bar \chi} \partial {\rm X} + M^2  {\bar \chi} {\rm X} \right. \\
\label{skeld}
&& \hspace{1in} \left.  +  g {\bar \chi} {\bar \varphi} \Phi + \frac{g}{2} {\rm X} \left( {\bar \varphi}^2 + \frac{  {\Phi}^2}{4} \right)  \right]  .
\end{eqnarray}
In this Keldysh basis the propagators are the advanced and retarded Green's functions $G^{A,R}$ and the Wightman function $F$:
\begin{eqnarray}
\label{keldyshgreens}
F_{\bf k}(\tau_1,\tau_2) &\equiv& \langle {\bar \varphi}_{\bf k}(\tau_1) {\bar \varphi}_{\bf -k}(\tau_2) \rangle \\
\nonumber
&=& {\rm Re} \left[ U_{\bf k}(\tau_1) U^*_{\bf k}( \tau_2) \right] ,  \\
\nonumber
G^R_{\bf k}(\tau_1,\tau_2) &\equiv& i \langle {\bar \varphi}_{\bf k} (\tau_1) \Phi_{\bf -k}(\tau_2) \rangle \\
\nonumber
&=& - 2 \theta(\tau_1-\tau_2) {\rm Im} \left[ U_{\bf k}(\tau_1) U^*_{\bf k}( \tau_2) \right] , \\
\nonumber
G^A_{\bf k}(\tau_1,\tau_2) &\equiv& G^R_{\bf k}(\tau_2,\tau_1), \\ \nonumber 
0 &=&\langle \Phi_{\bf k}(\tau_1) \Phi_{\bf -k}(\tau_2) \rangle
\end{eqnarray}
where 
\[ U_{\bf k}(\tau) = {\frac{H}{\sqrt{2 k^3}} \left( 1 - i k \tau \right) e^{-ik \tau}  } \]
is a solution to the free massless $\varphi$-equation of motion, chosen to obey Bunch Davies boundary conditions at early times.  For the massive $\chi$-field one has ${\cal F}_{\bf k}(\tau_1,\tau_2)={\rm Re}[V_{\bf k}(\tau_1)V^{\ast}_{\bf k}(\tau_2)]$, etc. where the free field solution $V_{\bf k}(\tau)$ can be approximated by
\begin{eqnarray*}
V_{\bf k} (\tau) 
&\approx& -\frac{H\tau\exp \left[ - i \int ^\tau _{\tau_{\rm in}} d \tau' \sqrt{  k^2 + \frac{M^2}{H^2 \tau'^2}} \right]}{\sqrt{ 2} \left( k^2 + \frac{M^2}{H^2 \tau^2} \right)^{1/4} } 
\end{eqnarray*}
in the WKB limit $| \dot{\omega}|/\omega^2 \ll 1$, which is always valid for $H/M \ll 1$.  

In the decoupling limit $g \rightarrow 0$ or $M \rightarrow \infty$, the inflaton fluctuation power spectrum is simply
\begin{equation}
\label{ps0}
P^{(0)}_{\varphi} = \frac{k^3}{2 \pi^2} F_{\bf k}(0,0) = \left( \frac{H}{2\pi} \right)^2. 
\end{equation}
Corrections to this will come from 
the interactions (\ref{skeld}), %KS Start
which contribute to
all connected diagrams with two external solid lines (Fig. 1).  We have assumed that all tadpoles (1-point diagrams) can be canceled via local counterterms, i.e. the cosmological background is quantum-mechanically stable; this issue is more thoroughly addressed in a forthcoming treatment of the details of  renormalization in cosmological backgrounds \cite{jacksonschalmvdaalst}.

Let us analyze the first diagram (recall that future infinity equals $\tau=0^-$),
\begin{eqnarray*}
 P^{({\rm A})}_{\varphi}(k) &=& \frac{k^3}{2 \pi^2} \left( -i g \right)^2  \int_{\tau_{\rm in}}^0 d \tau_1 \ a(\tau_1)^4 \int_{\tau_{\rm in}}^0 d \tau_2 \ a(\tau_2)^4 \times \\
&& \hspace{-0.9in} \int \frac{d ^3 \bf q}{(2 \pi)^3} \left[ -i G^R_{\bf k}(0,\tau_1) \right]  \mathcal F_{\bf q+k}(\tau_1,\tau_2) F_{\bf q}(\tau_1,\tau_2) \left[ -i G^A_{\bf k}(\tau_2,0) \right].
\end{eqnarray*}
Writing out the Green and Wightman functions in terms of $U$'s and $V$'s, we see there are three types of vertices.  The first is
%KS Start Added f(\tau) back
\begin{eqnarray*}
\mathcal A_1( {\bf k}_1,{\bf k}_2) &\equiv& \int _{\tau_{\rm in}}^0 d \tau \ a(\tau)^4 {U}_{{\bf k}_1}(\tau)  {U}_{{\bf k}_2}(\tau)  V^*_{-({\bf k}_1+{\bf k}_2)}(\tau) f(\tau) \\
&& \hspace{-0.5in} = - \frac{1}{2  \sqrt{2k_1^3 k_2^3} H } \int _{\tau_{\rm in}}^0 \frac{d \tau}{\tau^3}  \frac{  \left( 1-i k_1 \tau \right) \left( 1-i k_2 \tau \right) }{\left( | {{\bf k}_1+{\bf k}_2}|^2 + \frac{M^2}{H^2 \tau^2} \right)^{1/4} }f(\tau)  \\
&& \hspace{-0.8in} \times \exp \left[ - i (k_1 + k_2) \tau + i \int ^{\tau}_{\tau_{\rm in}} d \tau' \sqrt{ | {{\bf k}_1+{\bf k}_2}|^2 + \frac{M^2}{H^2 \tau'^2}} \right] ,
\end{eqnarray*}
where we have introduced a function $f(\tau)$ to account for any step-functions.
By introducing the rescaled time $u \equiv (H/M) \tau$, the vertex $\mathcal A_1( {\bf k}_1,{\bf k}_2)$ admits a stationary phase approximation at the energy-conservation moment
\begin{equation}
\label{statphase}
 k_1+ k_2 = \sqrt{  | {\bf k}_1+{\bf k}_2|^2 + u_c^{-2} } . 
 \end{equation}
The solution to this defines the New Physics Hypersurface (NPH),
\[ u_{c}^{-1}= - \sqrt{2k_1k_2(1-\cos\theta)}, \hspace{0.3in} \cos\theta=\frac{ {\bf k}_1 \cdot {\bf k}_2}{k_1 k_2}. \]
Then to leading order in $H/M$ the amplitude is
\begin{eqnarray}
\nonumber
\mathcal A_1( {\bf k}_1,{\bf k}_2)  &\approx& - \frac{ \sqrt{\pi i} f(\tau_c) e^{-i \frac{M}{H} \sqrt{ | {\bf k}_1 + {\bf k}_2|^2  u_{\rm in}^2+ 1} }}{ 2\sqrt{k_1 k_2} \left[ 2 k_1 k_2 (1- \cos \theta ) \right]^{1/4} \sqrt{ HM}} \\
\label{a1}
&&\hspace{-0.5in} \times  \left( \frac{ {k_1 + k_2 + \sqrt{2k_1 k_2 (1- \cos \theta)}}}{\sqrt{ | {{\bf k}_1+{\bf k}_2}|^2 +  u_{\rm in}^{-2} } + |u_{\rm in}|^{-1}} \right) ^{-i \frac{M}{H} } .
\end{eqnarray}
The physics of this is clear. This diagram accounts for the threshold production/decay of heavy particles at high redshift in the early universe.  Note that in order to evaluate $f(\tau_c)$, one should use the step-function appropriately ``averaged" due to the Gaussian fluctuations:
\begin{equation}
\label{avstep}
 \theta(\tau) = \left\{ 
\begin{array}{cc} 
1 & {\rm if} \ \tau > 0, \\ 
\sfrac{1}{2} & {\rm if} \ \tau = 0, \\ 
0  & {\rm if} \ \tau < 0. \\ 
\end{array} \right. 
\end{equation}

The second possible vertex is identical to $\mathcal A_1$ but with one $U$ conjugated.  This has only imaginary-time saddlepoint solutions.  Since our $\tau$-integral is confined to the real axis we will never pass over this point in our integration, and so this amplitude will be suppressed as $\mathcal A_2 \sim {\rm erf} (\frac{M}{H}) \sim \frac{H}{M} e^{-(M/H)^2}$, allowing us to neglect such interactions.  Finally we consider $\mathcal A_3$ which has both $U$'s conjugated and so admits no saddlepoint solutions, and thus can also be neglected.

\begin{figure}
\begin{center}
\parbox{40mm}{\includegraphics[scale=0.18]{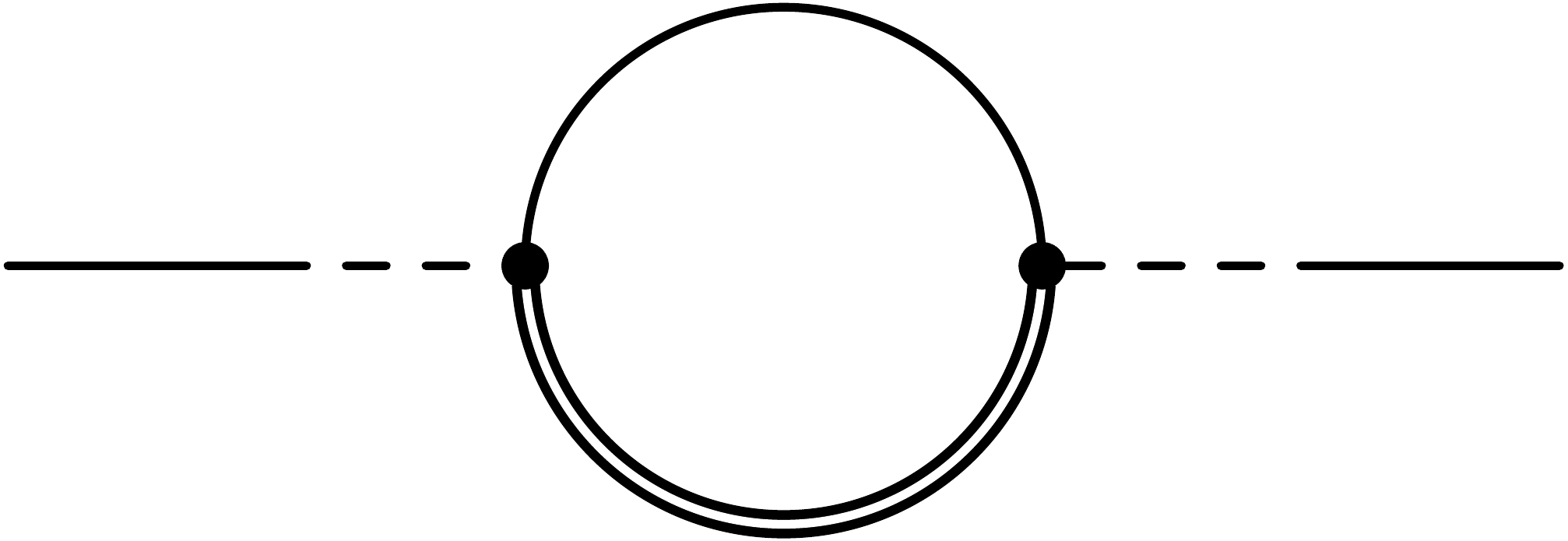} \hspace{-0.1in} \newline A}
\parbox{40mm}{\includegraphics[scale=0.18]{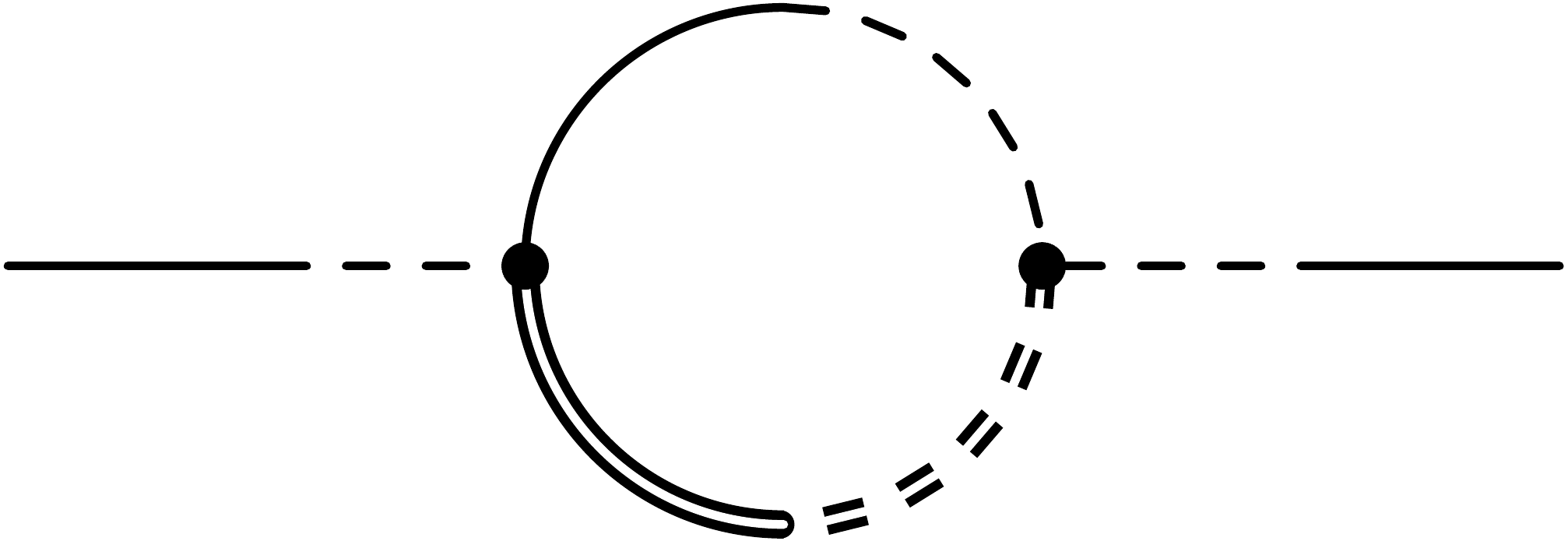} \hspace{-0.1in} \newline B} \\
\vspace{0.1in}
\parbox{40mm}{\includegraphics[scale=0.18]{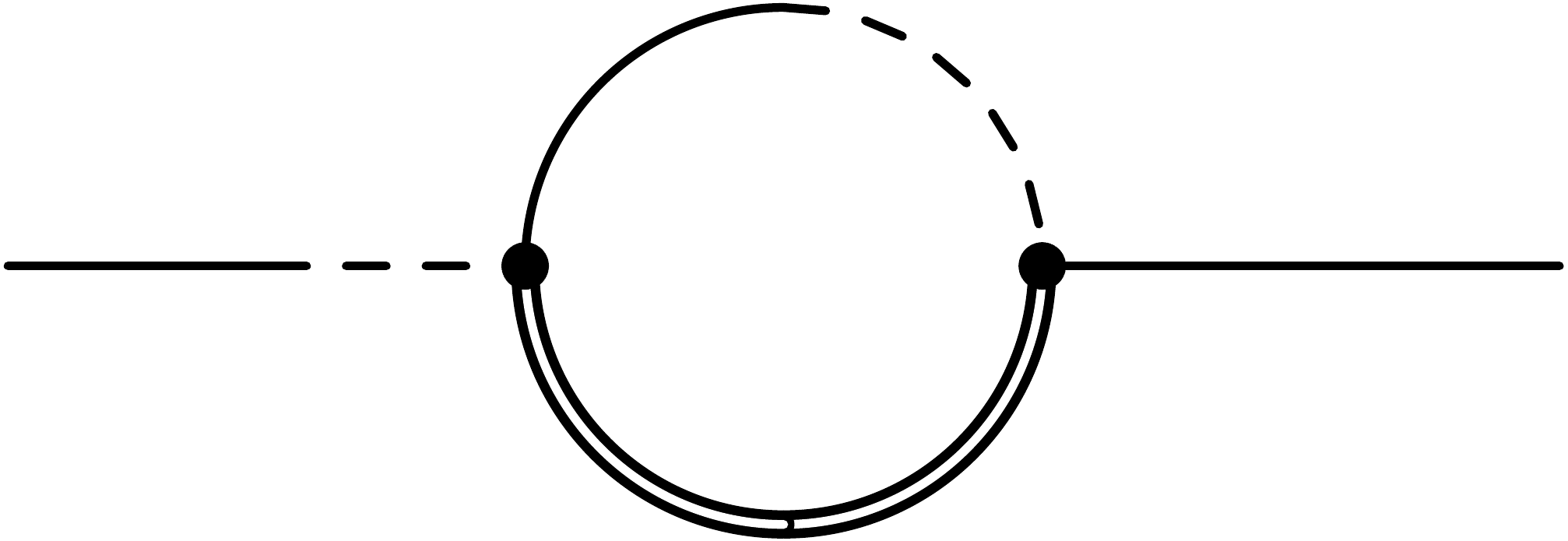} \hspace{-0.1in} \newline C} 
\parbox{40mm}{\includegraphics[scale=0.18]{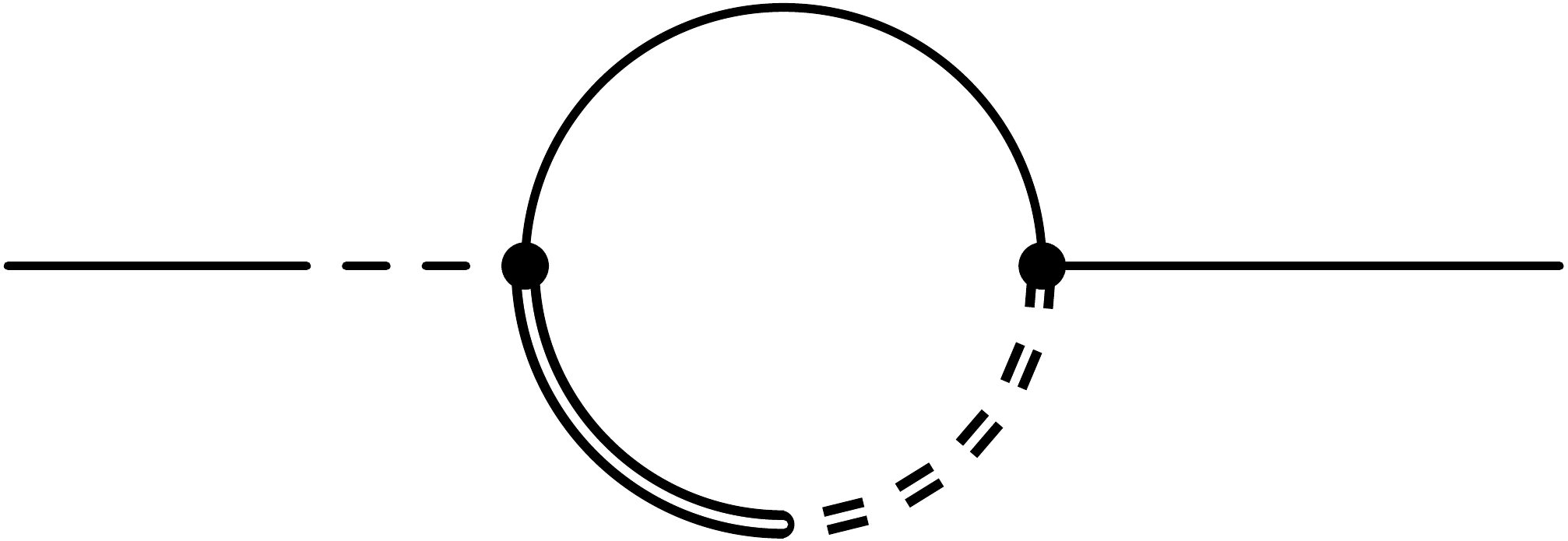} \hspace{-0.1in} \newline D} 
\caption{Power spectrum corrections mediated by the heavy field.  Single solid lines indicate contractions of ${\bar \varphi}$, dashed single lines indicate those of $\Phi$, with analogous notation for double lines indicating the heavy field components $\{ {\bar \chi},{\rm X} \}$.}
\label{boundarycorrections}
\end{center}
\end{figure}

The integrals over $\tau_1,\tau_2$ yield \eqref{a1} and its conjugate, removing any phase. The integral over the loop-momentum should then be traded for an integral over the NPH.  While one cannot truly assign an energy to a field in an expanding background, we can define an energy in the practical sense.  Multiply the stationary phase definition (\ref{statphase}) by $H |\tau_c|$ to yield the physical energy conservation at the NPH moment,
\[ E_q + E_k = E_\chi  \]
where $E_q$ is the energy of the virtual $\varphi$-field, $E_k$ is the energy of the external $\varphi$-field and $E_\chi$ is the energy of the $\chi$-field at the moment of interaction.  To evaluate the integral over ${\bf q}$, we perform the coordinate transformation given by
\begin{eqnarray}
\nonumber
\tau^{-1} &\equiv& - \frac{H}{M} \sqrt{2 kq(1-\cos \theta)}, \\
\label{qenergy}
E_q \equiv H q |\tau| &=& M \sqrt{ \frac{q}{2k(1-\cos \theta)} } . 
\end{eqnarray}
The ${\bf q}$-integral then transforms as
\begin{eqnarray}  
\nonumber
\hspace{-0.0in} \int \frac{d ^3 \bf q}{(2 \pi)^3} \left| \mathcal A_1({\bf k}, {\bf q} ) \right|^2 &=& \frac{1}{16 \sqrt{2} \pi k^{3/2} M} \int \frac{ \sqrt{q} dq d (1-\cos \theta)}{H \sqrt{1-\cos \theta}}\\
\label{eq:23}
&\rightarrow& - \frac{1}{16 \pi k^2 H^4} \int \frac{d\tau d E_q}{\tau^{3} } . 
\end{eqnarray}
Imposing the UV constraint $E_\chi \leq \Lambda$, and the geometrical constraint $1 - \cos \theta \leq 2$, one finds
% KS End
\begin{eqnarray*}
\frac{M^2}{4 H k |\tau|} \leq & E_q & \leq - k | \tau|H + \Lambda, \\
-\frac{\Lambda + \sqrt{\Lambda^2 - M^2}}{2 Hk} \leq &\tau& \leq -\frac{\Lambda - \sqrt{\Lambda^2 - M^2}}{2 Hk}.
\end{eqnarray*}
Performing the integrals, the leading $\Lambda \gg M$ correction for this diagram is then the scale-invariant result
\begin{equation}
\label{psa}
\Delta P^{({\rm A})}_{\varphi} \approx \frac{g^2 H \Lambda^3}{96 \pi^3 M^4} . 
\end{equation}

The second correction, represented by diagram B, can be evaluated in a similar manner but contains a minus sign relative to diagram A.  There is also a subtlety in that there is now a Heaviside function $\theta(\tau_1 - \tau_2)$ producing a factor of $\sfrac{1}{2}$ via eq. (\ref{avstep}),
\begin{equation}
\label{psa}
\Delta P^{({\rm B})}_{\varphi} \approx -\frac{g^2 H \Lambda^3}{192 \pi^3 M^4} . 
\end{equation}
Finally, it is easy to see that diagrams C and D cancel against each other to leading order: by ignoring the external lines, they can be seen to be corrections to the Green's function rather than initial state effects.  Thus to leading order in $H/M$,
\begin{equation}
\label{psa}
\Delta P_{\varphi} \approx \frac{g^2 H \Lambda^3}{192 \pi^3 M^4}
\end{equation}
is the complete power-spectrum correction due to high-energy physics.  It is a simple overall enhancement of the amplitude without any characteristic momentum-dependent features.  In a slow-roll background $H$ will acquire a weak dependence on $k$; the full details of the slow-roll expansion are presented a companion article \cite{Jackson:2011qg}.

%%%%%%%%%%%%%%%%%%%%%%%%%%%%
\subsection{Universal Effective Action, Vacuum Choice, Observability and Conclusion}
%%%%%%%%%%%%%%%%%%%%%%%%%%%%
Intuitively there exists a low energy effective action in terms of only the inflaton fluctuations $\varphi$  which reproduces these corrections.  First Fourier expand as
\[ {\bar \varphi}_{{\bf q}_i} (\tau) = \frac{1}{a(\tau)} \int \frac{d \omega_i}{2 \pi} {\tilde {\bar \varphi}}_{{\bf q}_i, \omega_i} e^{-i \omega_i \tau} \]
and similarly for $\Phi$.  
Integrating out $\chi$, the leading term in $H/M$ is \cite{Jackson:2011qg}:
\begin{eqnarray}
\nonumber
&& \hspace{-0.3in} \mathcal S_{\rm int,4} [ {\bar \varphi}, {\Phi}] = -
\int \prod_i \frac{d \omega_i d^3 {\bf q}_i}{(2\pi)^4} (2\pi)^3 \delta^3(\sum_i {\bf q}_i)  \\ 
%\label{new4pteff}
\nonumber
&& \hspace{-0.3in} \times \frac{g^2}{2!} \left( 2 {\tilde {\bar {\varphi}}}_{1}  {\tilde {\Phi}}_{2} \theta \left( \tau_{1c} - \tau_{2c} \right) \right. \\
\nonumber
&& \hspace{-0.3in}  \left. {\rm Im}   \left[  \mathcal B^*(\omega_1, \omega_2,{\bf q}_1 + {\bf q}_2) \mathcal B(\omega_3, \omega_4, {\bf q}_3 + {\bf q}_4) \right] \left( {\tilde {\bar{\varphi}}}_{3}  {\tilde {\bar {\varphi}}}_{4} + \frac{1}{4} {\tilde {{\Phi}}}_{3}  {\tilde {{\Phi}}}_{4} \right) \right. \\
\nonumber
&& \left. \hspace{-0.2in} + i {\tilde {\bar {\varphi}}}_{1}  {\tilde \Phi}_{2}  {\rm Re} \left[ \mathcal B^* ( \omega_1, \omega_2,{\bf q}_1 + {\bf q}_2) \mathcal B ( \omega_3, \omega_4, {\bf q}_3 + {\bf q}_4)\right] {\tilde {\bar{\varphi}}}_{3}  {\tilde {\Phi}}_{4}  \right) 
\end{eqnarray}
where
\begin{eqnarray*}
\mathcal B (\omega_1, \omega_2, {\bf q}) &=& \int^0_{\tau_{\rm in}} d \tau \ a(\tau)^2 e^{-i (\omega_1+\omega_2) \tau} V^*_{\bf q}(\tau) \\
&=& \frac{1}{\sqrt{2} } \int^0_{\tau_{\rm in}} \frac{d \tau}{ H \tau}  \frac{ e^{-i  (\omega_1 + \omega_2) \tau +i\int ^{\tau}_{\tau_{\rm in}} d \tau' \sqrt{ q^2 + \frac{M^2}{H^2 \tau'^2}}  } }{\left( q^2 + \frac{M^2}{H^2 \tau^2} \right)^{1/4} }
\end{eqnarray*}
which can be evaluated using a stationary phase approximation, and
\begin{eqnarray*}
\tau^{-1}_{1c} &=& - \frac{H}{M} \sqrt{ | \omega_1 + \omega_2|^2 - | {\bf q}_1 + {\bf q}_2|^2} , \\
\tau^{-1}_{2c} &=& - \frac{H}{M} \sqrt{ | \omega_3 + \omega_4|^2 - | {\bf q}_3 + {\bf q}_4|^2} .
\end{eqnarray*}
%KS New

We see that this effective action has specically localized interactions on the NPH. It therefore has the virtues of both the NPH and BEFT models without either of their vices.  As a ``generalized'' boundary effective action it can be connected to microscopic physics, but it is controlled by the NPH.  

Our computation reveals that the leading universal generic contribution to the inflationary power spectrum is indeed unambiguously of linear order in $H/M$.
On the other hand its profile, a flat enhancement, is qualitatively different from what was surmised. Both initial state NPH and BEFT approaches indicated a characteristic oscillatory signal in the generic correction due to initial states \cite{Greene:2005aj} With the fully consistent approach to compute the universal generic correction pioneered, we now can trace the origin of this oscillatory behavior. If one would chose a co-moving cut-off instead of a physical cut-off as in eq.~\eqref{eq:23}, one cannot make tadpoles vanish consistently. The remaining terms yield the oscillatory signal. A first draft of this article showed this explicitly. Since the presence or absence of oscillatory features depends on the cut-off used, it cannot be a physical effect and in should be absent in a properly renormalized theory when the cut-off is removed after the introduction of counterterms \cite{jacksonschalmvdaalst}. This does not mean that one can never have oscillatory features in the inflationary power spectrum. It is just that they are not a generic prediction of unknown high energy physics, but rather of some non-generic phenomenon, e.g. \cite{Adams:2001vc,Kaloper:2003nv,Achucarro:2010da,Chen:2011zf}.

% KS Drop below
% We could then contract a set of light fields in this quartic-field effective action to produce the quadratic-field effective action.  This is dominated by the induced density matrix of the initial state, $\ln \rho \sim - \Gamma {\tilde \Phi}^2$, allowing small departures from on-shell coherence.  It does not appear possible to produce specific departures from the Bunch-Davies vacuum, however.

% The vacuum could potentially be modified as follows.  The stationary phase condition (\ref{statphase}) yields solutions which come in pairs: the expected one we employed for which $\tau_c < 0$, as well as another one for which $\tau_c > 0$.  Although we have limited ourselves to $\tau_c \in (-\infty, 0^-)$, the $\tau_c > 0$ solution is perfectly valid and corresponds to the \emph{contracting} phase of de Sitter space, though this is not usually considered in standard inflationary theory.  If we allow our background to include such a contracting phase, there would be additional effective interactions induced on the boundary at $\tau = \pm \infty$, corresponding to initial condition selection.

% \begin{figure}
% \begin{center}
% \includegraphics[width=3in]{pk.pdf}
% \caption{Power spectra for the example theory (red), and the free theory (black).  Slight scale-dependence will be induced from slow-roll variation of the Hubble parameter $H$.}
% \label{pk}
% \end{center}
% \end{figure}
% KS End Drop

In summary, we have developed a technique to explicitly calculate {\em universal generic } corrections to the inflaton power spectrum from fundamental high energy physics.  While the contribution from each loop-momentum is localized to a unique New Physics Hypersurface \cite{Easther:2001fi}, the integral over such loop-momenta yields a correction to which is widely distributed in time.  The result is a scale-invariant overall enhancement at order $H/M$ to the power spectrum.  This allows us to effectively represent microscopic models through a ``generalized'' Boundary Effective Field Theory as in \cite{Schalm:2004qk}.  This effective action includes a spreading of the initial state density matrix, producing a loss of quantum coherence.  Most importantly, these corrections are potentially observable;
a definitive statement on this requires further detailed study. 

{\bf Acknowledgments:}
We would like to thank \mbox{U. Danielsson}, R. Easther, B. Greene, W. Kinney, \mbox{M. Kleban}, L. McAllister, M. Parikh, G. Shiu, and \mbox{J. P. van der Schaar} for discussions past and present.  This research
was supported in part by a VIDI and a VICI Innovative Research
Incentive Award from the Netherlands Organisation for Scientific
Research (NWO), a van Gogh grant  from the NWO, and the Dutch
Foundation for Fundamental Research on Matter (FOM).

\end{document}